# Magnetic Antiskyrmions in Two-Dimensional van der Waals Magnets Engineered by Layer Stacking

Kai Huang,[1] Edward Schwartz,[1] Ding-Fu Shao,[2] Alexey A. Kovalev,[1] and Evgeny Y. Tsymbal[1]

[1] Department of Physics and Astronomy & Nebraska Center for Materials and Nanoscience, University of Nebraska, Lincoln, Nebraska 68588-0299, USA

[2] Key Laboratory of Materials Physics, Institute of Solid-State Physics, HFIPS, Chinese Academy of Sciences, Hefei 230031, China

**ABSTRACT:** Magnetic skyrmions and antiskyrmions are topologically protected quasiparticles exhibiting a whirling spin texture in real space. Antiskyrmions offer some advantages over skyrmions as they are expected to have higher stability and can be electrically driven with no transverse motion. However, unlike the widely investigated skyrmions, antiskyrmions are rarely observed due to the required anisotropic Dzyaloshinskii-Moriya interaction (DMI). Here we propose to exploit the recently demonstrated van der Waals (vdW) assembly of two-dimensional (2D) materials that breaks inversion symmetry and creates conditions for anisotropic DMI. Using a 2D vdW magnet $CrI_3$ as an example, we demonstrate, based on density functional theory (DFT) calculations, that this strategy is a promising platform to realize antiskyrmions. Polar layer stacking of two centrosymmetric magnetic monolayers of $CrI_3$ efficiently lowers the symmetry, resulting in anisotropic DMI that supports antiskyrmions. The DMI is reversible by switching the ferroelectric polarization inherited from the polar layer stacking, offering the control of antiskyrmions by an electric field. Furthermore, we find that the magnetocrystalline anisotropy and DMI of $CrI_3$ can be efficiently modulated by Mn doping, creating a possibility to control the size of antiskyrmions. Using atomistic spin dynamics simulations with the parameters obtained from our DFT calculations, we predict the formation of antiskyrmions in a $Cr_{0.88}Mn_{0.12}I_3$ bilayer and switching their spin texture with polarization reversal. Our results open a new direction to generate and control magnetic antiskyrmions in 2D vdW magnetic systems.

## INTRODUCTION

The discovery of magnetic skyrmions [1-6] has led to an increasing interest in magnetic quasiparticles exhibiting a whirling spin texture and non-trivial real space topology [7-9]. The topological protection of these quasiparticles ensures their shape stability and makes them promising as potential information carriers in spintronic devices [10-12]. The topology of the magnetic quasiparticles is characterized by a topological charge [13], a nonvanishing integer describing the winding number of the classical spin when mapped on a unit sphere. The topological charge is associated with interesting physical phenomena, such as the topological Hall effect [14,15] and the skyrmion Hall effect [16-18] that can be used to detect the quasiparticles.

Among the topological quasiparticles are magnetic antiskyrmions. Antiskyrmions are analogs of skyrmions but have an opposite topological charge compared to a skyrmion with the same polarity [19]. The morphological difference between skyrmions and antiskyrmions comes from their symmetries. It is known that depending on the crystal symmetry, two distinct types of skyrmions can be observed, Bloch [2, 3] and Néel skyrmions [20]. Skyrmions, no matter Bloch type or Néel type, are always isotropic. As shown in Fig. 1(a,c), the spin spiral structure along $x$ and $y$ directions of a Néel-type skyrmion is the same. However, in antiskyrmions, the in-plane rotational symmetry is broken and the spin structures along the $x$ and $y$ directions have opposite chiralities and contain alternating Bloch- and Néel-type spin twists [19,21] (Fig. 1(b,d)).

Skyrmions and antiskyrmions are typically stabilized by Dzyaloshinskii-Moriya interaction (DMI) – an antisymmetric exchange interaction between adjacent magnetic moments that favors their perpendicular alignment [22,23]. The DMI between spins $\vec{S}_\alpha$ and $\vec{S}_\beta$ takes the form

$$E_{\text{DMI}} = \vec{d}_{\alpha\beta} \cdot (\vec{S}_\alpha \times \vec{S}_\beta), \qquad (1)$$

where $\vec{d}_{\alpha\beta}$ is the DMI vector. The DMI energy is controlled by symmetry of the system and spin-orbit coupling. It requires broken inversion symmetry (which is a necessary but not sufficient condition) and can be isotropic or anisotropic. In thin-film structures the isotropic DMI leads to skyrmions (Fig. 1(a,c)), while the anisotropic DMI with opposite signs along perpendicular $x$ and $y$ directions can stabilize antiskyrmions (Fig. 1(b,d)).

The requirement of the anisotropic DMI makes observing magnetic antiskyrmions more difficult. Nevertheless, recently, there have been a few experimental demonstrations of antiskyrmions in noncentrosymmetric magnets of $D_{2d}$ [24-26] and $S_4$ [27] point-group symmetries. Theoretical analyses indicate that antiskyrmions may have some advantages over their isotropic counterparts. In particular, antiskyrmions are expected to have stronger stability due to the reduced magnetostatic energy [28]. In addition, antiskyrmions can be driven by charge currents without a transverse motion due to the skyrmion Hall effect, which may be beneficial from the point of view of applications in race-track memories [19,29].

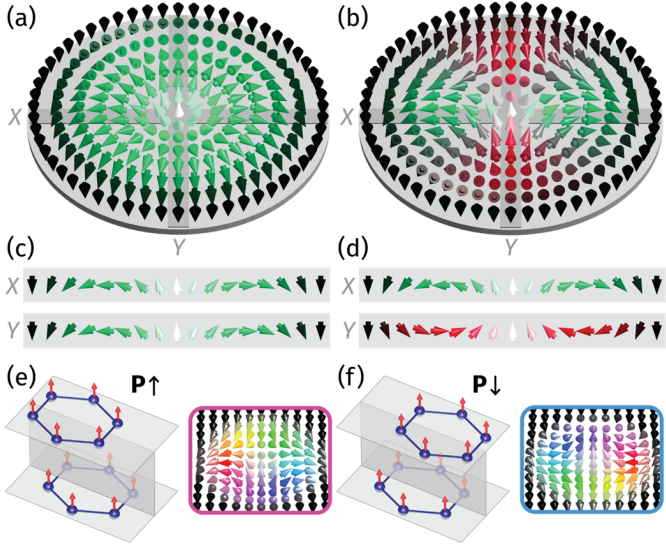

**Figure 1** (**a**, **b**) Schematic of a Néel-type skyrmion (a) and anti-skyrmion (b). Brightness of each arrow represents an out-of-plane spin component with a bright (dark) contrast indicating spin pointing up (down). Color represents the chirality of spin rotation along the diameter in the specific direction, where green (red) color indicates clockwise (counterclockwise) rotation. (**c**, **d**) Spin rotation of the skyrmion (c) and antiskyrmion (d) along the diameter in $x$ and $y$ directions. (**e**, **f**) Schematic of 2D ferromagnetic bilayers engineered to have polar stackings with ferroelectric polarization pointing up $P\uparrow$ (e) or down $P\downarrow$ (f). Red arrows indicate magnetic moments. Polarization switching changes the spin texture of magnetic antiskyrmions.

Two-dimensional (2D) van der Waals (vdW) ferromagnets [30 - 33] have recently emerged as novel skyrmionic hosts. In these materials, which are all centrosymmetric, space inversion symmetry breaking required for DMI can be realized by the interface proximity effect in a vdW heterostructure [34, 35]. This allows the stabilization of magnetic skyrmions in a system with minimum thickness and even in the absence of the external magnetic field, as has been demonstrated in the recent experiments [36-40].

Besides the interface proximity effect, recent advances in vdW assembly techniques showed that interlayer stacking can be used to control properties of vdW materials. It has been demonstrated that minor interlayer sliding [41,42] or small-angle interlayer twisting [43-45] can generate electronic and transport responses of 2D materials which do not exist in the bulk-like phase. Specifically, a combination of 180° rotation of the top monolayer followed by interlayer sliding results in a non-centrosymmetric stacking pattern (Fig. 1 (e,f)). This is accompanied by the emergence of an out-of-plane electric polarization reversible by an external electric field through interlayer sliding [46-48]. This approach allows the design of synthetic 2D ferroelectrics (and multiferroics) out of parent nonpolar compounds and uncover new functionalities not existent in the bulk phase (e.g., [49]).

Applying this approach to 2D vdW ferromagnets allows breaking inversion symmetry and thus providing the prerequisite for DMI. In addition, sliding away from the mirror-symmetric stacking to a more energetically favorable stacking breaks the in-plane rotational symmetry of the 2D crystal producing anisotropy of the DMI that is required for antiskyrmions. The switching of ferroelectric polarization by an applied electric field through interlayer sliding is expected to reverse the DMI sign, providing a possibility of a non-volatile electric-field control of antiskyrmions.

Among 2D vdW magnets, $CrI_3$ appears to be a viable material to exhibit antiskyrmions when it is engineered to be polar. $CrI_3$ has been one of the first vdW magnets exfoliated down to a monolayer [31] and extensively studied afterwards [50-52]. It has out-of-plane magnetic anisotropy required for magnetic (anti)skyrmions. The heavy I atoms are expected to provide a large DMI when symmetry allows. In fact, theoretical calculations have predicted the formation of magnetic skyrmions in a $CrI_3$ monolayer in the presence of a strong external electric field [53] or defects [54]. Lastly, $CrI_3$ is an insulator, which allows applying a sufficiently large electric field across a polar $CrI_3$ bilayer to switch its ferroelectric polarization.

In this work, using first-principles density functional theory (DFT) calculations, we explore the possibility to realize magnetic antiskyrmions in a ferroelectric $CrI_3$ bilayer engineered by layer stacking. We demonstrate that the polar layer stacking leads to anisotropic DMI, that can be switched by reversal of ferroelectric polarization. Further, we show that Mn doping of $CrI_3$ is favorable for magnetic antiskyrmions to occur in a polar $Cr_{1-x}Mn_xI_3$ bilayer due to the reduced magnetic anisotropy and enhanced DMI. Our atomistic spin dynamics simulations with the parameters obtained from the DFT calculations predict the formation of antiskyrmions in a $Cr_{0.88}Mn_{0.12}I_3$ bilayer and switching their spin texture with polarization reversal. These results demonstrate a new approach to generate magnetic quasiparticles in 2D vdW magnets and control their spin texture by electric fields.

## RESULTS

**Symmetry Analysis.** $CrI_3$ is a 2D vdW ferromagnetic insulator with an easy axis along the [001] direction ($z$-axis). At low temperature, bulk $CrI_3$ is stabilized in a rhombohedral structure of space group $R\bar{3}$ [55]. The bulk unit cell contains three $CrI_3$ monolayers stacked in the [001] direction with Cr atoms surrounded by a distorted edge-sharing octahedron of six I atoms (Fig. 2(a)). The Cr-I-Cr bond angle between the $CrI_6$ octahedra is nearly 90°, leading to the superexchange interaction that favors a ferromagnetic intralayer coupling [56,57]. The presence of inversion symmetry in bulk $CrI_3$ prohibits a finite DMI, and hence magnetic quasiparticles are not expected to emerge in bulk $CrI_3$. The same conclusion is valid for monolayer $CrI_3$ whose structure belongs to centrosymmetric magnetic space group $P\bar{3}1m$ and thus prohibits DMI. Therefore, an interface proximity effect or an external electric field [53] are typically considered as means to break inversion symmetry and create conditions for the appearance of topological magnetic quasiparticles.

Here, we explore a different approach where the polar stacking technique is used to break inversion symmetry. First, a $CrI_3$ bilayer is formed in such a way that the top monolayer $\bar{A}$ is assembled as a mirror reflection $M_z$ of the bottom monolayer A with respect to the (001) plane (Fig. 2(b)). This stacking, denoted by A$\bar{\text{A}}$, is related to the bulk-like stacking of a CrI$_3$ bilayer by 180° rotation of the top monolayer about the $z$ axis. While this staking breaks the inversion symmetry, it does not produce DMI in the 2D system due to $M_z$ symmetry. The A$\bar{\text{A}}$ structure is, however, usually energetically unfavorable. It spontaneously relaxes to a more favorable A$\bar{\text{B}}$ (or $\bar{\text{B}}$A) stacking through translation of the top monolayer layer along the [100] direction ($x$ axis) by $\pm a/3$, where $a$ is a lattice constant (Fig. 2(c,d)). Under this translation, the magnetic space group of the bilayer is lowered from $P\bar{3}1m$ to $Cm'$. The translation along the $x$ axis not only breaks mirror symmetry $M_z$, but also breaks in-plane rotational $C_3$ symmetry that makes DMI anisotropic.

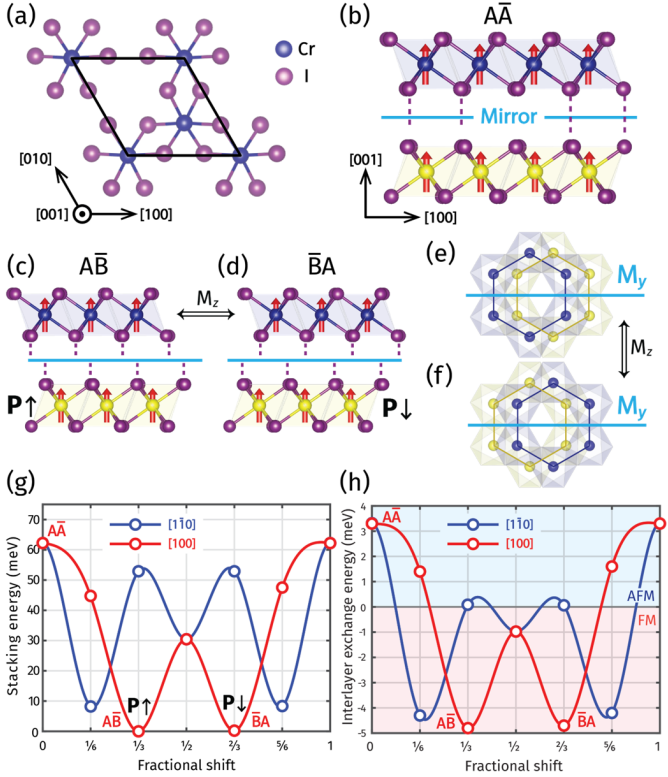

**Figure 2** (**a**) Top view of monolayer CrI$_3$. (**b**) Side view of A$\bar{\text{A}}$ stacking configuration of bilayer CrI$_3$ constructed by mirroring ($M_z$) of the bottom monolayer (yellow) to top (blue). (**c,d**) A$\bar{\text{B}}$ (c) and $\bar{\text{B}}$A (d) stacking configurations of bilayer CrI$_3$ obtained from A$\bar{\text{A}}$ by $\mp a/3$ translation of the top monolayer. They correspond to opposite ferroelectric polarizations ($P\uparrow$ and $P\downarrow$) and can be swapped by $M_z$ symmetry transformation. (**e, f**) Top views of (c,d), respectively. Only the Cr honeycomb lattice is shown. (**g, h**) Stacking energy (g) and the interlayer exchange energy (h) as a function of lateral translation along [100] (blue) and [1$\bar{1}$0] (red) directions with respect to the A$\bar{\text{A}}$ stacking (b).

To analyze the layer stacking effect on DMI and the spin texture, it is convenient to express the DMI energy in the form of a continuous model [58,59]

$$E_{\text{DMI}} = \sum_{ij} D_{ij}\vec{e}_i \cdot (\vec{m} \times \nabla_j \vec{m}), \quad (2)$$

where $D_{ij}$ is the DMI tensor with indexes $i, j \in x, y, z$, $\hat{e}_i$ is a unit vector, and $\vec{m}$ is the unit vector along the magnetization direction. The form of the DMI tensor $D_{ij}$ is determined by the crystallographic symmetry of the system. The CrI$_3$ monolayer structure belongs to point group $D_{3h}$, where all $D_{ij}$ components must be zero due to space inversion symmetry, resulting in a trivial spin texture. The same is true for the A$\bar{\text{A}}$ stacked bilayer due to $M_z$ symmetry. In the presence of an external electric field applied in the $z$ direction, the $D_{3h}$ symmetry of a CrI$_3$ monolayer is reduced to $C_{3v}$, which allows for two non-zero tensor components related by $D_{xy} = -D_{yx}$. Such DMI can stabilize Néel-type skyrmions [53]. The symmetry is further reduced for the polar A$\bar{\text{B}}$ ($\bar{\text{B}}$A) stacking, where the only remaining symmetry operation is mirror reflection $M_y$ (Fig 2. (e,f)). This symmetry permits the non-diagonal $D_{ij}$ components $D_{xy}, D_{yx}, D_{yz}$, and $D_{zy}$ to be non-zero and mutually independent (see Supporting Information for details and ref. 60 for a more general discussion). For a 2D system, however, we can assume that there is no magnetic moment variation along the $z$ direction resulting in $D_{yz} = 0$. Also, our calculations predict that $D_{zy}$ is small compared to $D_{xy}$ and $D_{yx}$ and therefore we ignore its contribution for our further analysis. These assumptions restrict the summation in Eq. (2) to $i, j \in x, y$.

According to Eq. (2), each $D_{ij}$ component corresponds to a spin spiral state along $j$ direction in the 2D system and the chirality of the spiral is defined by the sign of $D_{ij}$. Opposite sign of $D_{xy}$ and $D_{yx}$ favors the same spin spiral chirality along the $x$ and $y$ directions resulting in a skyrmion (Fig. 1 (a,c)), while the same sign of $D_{xy}$ and $D_{yx}$ favors opposite spin spiral chirality along the $x$ and $y$ directions resulting in an antiskyrmion (Fig. 1 (b,d)). If $|D_{xy}| = |D_{yx}|$, both skyrmions and antiskyrmions are expected to be circular, while if $|D_{xy}| \neq |D_{yx}|$, they can become elliptical. Elliptical skyrmions have been experimentally observed in systems with anisotropic DMI (e.g., [61]).

**DFT Results for a CrI$_3$ Bilayer.** Our DFT calculations are performed as described in Supporting Information. First, we ascertain that the ground state of bulk CrI$_3$ is predicted correctly. We find the magnetic moment of Cr atoms in the rhombohedral low-temperature bulk state is $m(\text{Cr}) = 3.32\ \mu_\text{B}$. The calculated intralayer Heisenberg exchange interaction is $J_{\text{intra}} = 1.81$ meV, supporting ferromagnetic ordering in the plane of CrI$_3$. The magnetic anisotropy energy (MAE) per Cr atom is found to be 0.72 meV. All these results are consistent with the previous calculations [62-65].

A polar bilayer is constructed by mirroring the bottom monolayer CrI$_3$ to the top. To find the ground state of the bilayer, we calculate the total energy as a function of the lateral alignment of the two monolayers. For each stacking configuration, the atomic structure of the bilayer is relaxed. As shown in Figure 2(g), two energy minima appear when the top monolayer is translated along the [100] direction with respect to the bottom monolayer, and three – when it is translated along the [1$\bar{1}$0] direction. As is evident from this plot, the $M_z$-symmetric A$\bar{\text{A}}$ stacking is unstable with respect to the lateral displacements. Two global energy minima appear at the $a/3$ and 2$a/3$ translations along the [100] direction ($\pm a/3$, when periodic boundary conditions are applied), corresponding to the $A\bar{B}$ and $\bar{B}A$ stacking configurations. $M_z$ symmetry is broken for these structures which makes them polar with the $A\bar{B}$ stacking having polarization up ($P\uparrow$, i.e. along the +z direction) and the $\bar{B}A$ stacking having polarization down ($P\downarrow$, i.e. along the −z direction. The two polar ground-state structures, $A\bar{B}$ and $\bar{B}A$, are related by $M_z$-symmetry transformation. Therefore, polarization switching of the bilayer is equivalent to the $\pm a/3$ lateral translation of one monolayer with respect to the other along the [100] direction (or equivalently along the [010] direction). The corresponding dipole moment per area for the polar structures is estimated to be 0.176 pC/m.

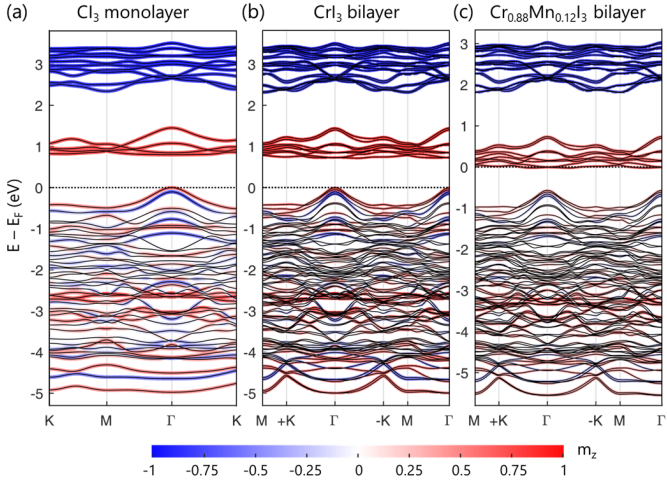

**Figure 3** (**a**, **b, c**) Band structure of a CrI$_3$ monolayer (a), a CrI$_3$ polar bilayer (b) and a Cr$_{0.88}$Mn$_{0.12}$I$_3$ polar bilayer (c) calculated in the presence of spin-orbit coupling. Spin projection to the direction of magnetization $m_z$ is shown in color. The horizontal dashed line indicates the Fermi energy ($E_F$).

We find that the intralayer exchange coupling in a CrI$_3$ bilayer is insensitive to the lateral alignment of the two monolayers favoring a ferromagnetic order. This is however not the case for the interlayer coupling which, due to its small magnitude, strongly depends on the layer stacking. Figure 2(h) shows the results of our computations. It is seen that while the $A\bar{A}$ stacking favors antiferromagnetic coupling with the Heisenberg exchange energy of about 3 meV per Cr atom, the polar $A\bar{B}$ and $\bar{B}A$ ground-state stacking exhibits ferromagnetic coupling of about 5 meV per Cr atom. These results are reminiscent to those found previously for a centrosymmetric CrI$_3$ bilayer [66] where the stacking-dependent magnetism was explained by a competition of the different interlayer orbital hybridizations. This fact indicates that while the polar stacking is essential for the appearance of non-vanishing DMI, it does not affect much the intralayer exchange coupling. Based on these results, in the following discussion, we assume that a polar $A\bar{B}$ ($\bar{B}A$)-stacked CrI$_3$ bilayer has all ferromagnetic order.

Figure 3(b) shows the calculated band structure of a polar bilayer CrI$_3$. Due to a relatively weak orbital hybridization across the interface, the electronic bands of the bilayer are similar to those of a monolayer (Fig. 1(a)). There are, however, visible differences between the bands around the +K and –K points for the bilayer but not for the monolayer. This is due to the broken space inversion symmetry of the bilayer that lifts band degeneracy for positive and negative wave vectors. Overall, the band structure of the bilayer is largely determined by the crystal field splitting of the 3$d$ orbitals of the Cr$^{3+}$ ion into the triply degenerate $t_{2g}$ states that lie at a lower energy and the doubly degenerate $e_g$ states that lie at a higher energy (bands around 1 eV in Fig. 3(a,b)). According to the Hund's rule, the majority-spin $t_{2g}$ states are fully occupied by the three available Cr$^{+3}$ electrons and form the valence band, while the majority-spin $e_g$ states form the low-energy conduction band. This picture is somewhat altered by the hybridization between the majority-and minority-spin bands driven by spin-orbit interaction, as well as by the Cr-I bonding that is seen from the orbital-resolved DOS in Figure S1. Due to a large exchange splitting of the Cr spin bands, the minority-spin $t_{2g}$ and $e_g$ states are empty and lie at a higher energy (bands around 3 eV in Fig. 3(a,b)).

Table 1 displays the calculated magnetic parameters for the polar CrI$_3$ bilayer in comparison to those obtained for bulk CrI$_3$. We find that the magnetic moment of Cr atom, the intralayer exchange coupling, and magnetocrystalline anisotropy do not change much by polar stacking. This is due to a weak interaction between CrI$_3$ monolayers so that stacking order doesn't influence the magnetic properties much. The critical difference occurs, however, in the DMI which changes from a zero value in bulk CrI$_3$ to non-zero in a polar bilayer.

**Table 1** Magnetic parameters of bulk CrI$_3$, polar bilayer CrI$_3$ ($P\uparrow$) and 12% Mn doped CrI$_3$ ($P\uparrow$): magnetic moment ($m$), intralayer exchange coupling ($J_{\text{intra}}$), magnetic stiffness ($A$), magnetocrystalline anisotropy ($K$), absolute values of DMI vectors ($|\vec{d}_1|$ and $|\vec{d}_2|$), and DMI tensor ($D$).

| Parameter | CrI$_3$ (bulk) | CrI$_3$ (polar bilayer) | Cr$_{0.88}$Mn$_{0.12}$I$_3$ (polar bilayer) |
|---|---|---|---|
| $m$ ($\mu_B$) | 3.32 | 3.31 | 3.43 (Cr) |
| | | | 3.66 (Mn) |
| $J_{\text{intra}}$ (meV) | 1.81 | 1.70 | 2.05 |
| $A$ (pJ/m) | 0.69 | 0.63 | 0.84 |
| $K$ (meV) | 0.72 | 0.67 | 0.03 |
| $K$ (MJ/m$^3$) | 0.84 | 0.77 | 0.04 |
| $|\vec{d}_1|$ (meV) | 0 | 0.04 | 0.08 |
| $|\vec{d}_2|$ (meV) | 0 | 0.05 | 0.13 |
| $D$ (mJ/m$^2$) | 0 | $\begin{bmatrix} 0 & -0.003 \\ -0.009 & 0 \end{bmatrix}$ | $\begin{bmatrix} 0 & -0.067 \\ -0.087 & 0 \end{bmatrix}$ |

Here, the calculation of DMI has been performed based on the following arguments. Microscopically, DMI arises from the interaction of two magnetic atoms mediated by a nonmagnetic atom. As a result, the DMI vector in Eq. (1), is given by $\vec{d}_{\alpha\beta} = d_{\alpha\beta}(\hat{r}_\alpha \times \hat{r}_\beta)$, where $\hat{r}_\alpha$ and $\hat{r}_\beta$ are the unit bond vectors between a nonmagnetic atom and magnetic atoms $\alpha$ and $\beta$, respectively. In our case, two magnetic Cr atoms are coupled through a nonmagnetic I atom via superexchange interaction. Based on the $Cm'$ symmetry of the system, we find two independent DMI vectors, $\vec{d}_1$ and $\vec{d}_2$, that control the DMI between adjacent Cr atoms (Supporting Information). For a CrI$_3$ monolayer, each of these DMI vectors can be viewed as the sum of two contributions arising from a superexchange interaction between two Cr atoms and top and bottom I atoms. For example, in Fig. 4(a), the vector $\vec{d}_1$ characterizing DMI between Cr$_0$ and Cr$_1$ magnetic moments is given by the sum of vectors $\vec{d}_{1a}$ and $\vec{d}_{1b}$ associated with the superexchange interaction of the Cr atoms via top atom I$_a$ and bottom atom I$_b$, respectively. Similarly, the DMI vector $\vec{d}_2$ that couples Cr$_0$ and Cr$_2$ moments, is given by $\vec{d}_2 = \vec{d}_{2a} + \vec{d}_{2b}$. In Figure 4(b), we plot all $\vec{d}_1$ and $\vec{d}_2$ vectors associated with each pair of Cr atoms in a Cr hexagon according to symmetry. It is seen that the $\vec{d}_1$ vectors (shown in red) lie off the mirror plane $M_y$ while the $\vec{d}_2$ vectors (shown in yellow) lie in that plane.

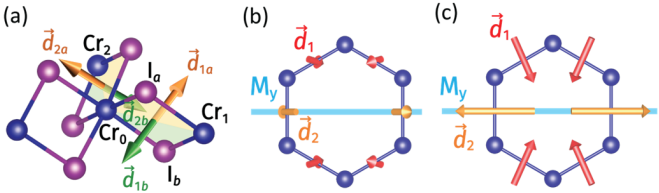

**Figure 4** (**a**) Construction of DMI vectors. DMI between Cr$_0$ and Cr$_1$ magnetic moments is determined by vector $\vec{d}_1$ given by the sum of vectors $\vec{d}_{1a}$ and $\vec{d}_{1b}$ resulting from the superexchange interaction of the Cr atoms with top atom I$_a$ and bottom atom I$_b$. Similarly, the DMI vector $\vec{d}_2$ that couples Cr$_0$ and Cr$_2$ moments is given by $\vec{d}_2 = \vec{d}_{2a} + \vec{d}_{2b}$. (**b, c**) DMI vectors $\vec{d}_1$ and $\vec{d}_2$ associated with pairs of Cr atoms connected by grey lines on a Cr honeycomb lattice for CrI$_3$ (a) and Cr$_{0.88}$Mn$_{0.12}$I$_3$ (b) polar bilayers. The DMI vectors in (b) and (c) are depicted in the same scale.

The DMI vectors are calculated using the spin-spiral method, where each component of $\vec{d}_1$ and $\vec{d}_2$ is obtained from the energy difference between specific spin configurations with a clockwise (CW) and counterclockwise (CCW) spiral. For example, as shown in Supporting Information,

$$d_{1x} = \frac{E_{\text{CCW}} - E_{\text{CW}}}{32S^2}, \quad (3)$$

where $S$ is the spin momentum. In the calculations, we construct the spin spirals assuming for simplicity that two monolayers of CrI$_3$ in a bilayer can be treated as one, which implies averaging of $\vec{d}_1$ and $\vec{d}_2$ over the two monolayers. The calculated DMI vectors $\vec{d}_1$ and $\vec{d}_2$ are depicted in Figure 4(b) and their values are listed in Table 1. The results show, as expected, that the in-plane rotational symmetry of the DMI in a bilayer is broken. As seen from Figure 4(b), the four DMI vectors $\vec{d}_1$ at sides of the Cr atom hexagon are centripetal, while the two DMI vectors $\vec{d}_2$ are centrifugal.

The obtained DMI vectors are related to the anisotropic DMI tensor in Eq. (2) as follows:

$$D_{ij} = \frac{1}{V_{\text{uc}}} \sum_{\alpha,\beta} S^2 d_{\alpha\beta,i} r_{\alpha\beta,j}, \quad (4)$$

where $\alpha, \beta$ represents different atoms in the unit cell, $V_{\text{uc}}$ is the volume of the unit cell, $d_{\alpha\beta,i}$ is the $i$ component of the DMI vector between atom $\alpha$ and $\beta$, $r_{\alpha\beta,j}$ is the $j$ component of the displacement vector from atom $\alpha$ to $\beta$. Using Eq. (1), we obtain two non-vanishing components of the DMI tensor: $D_{xy}$ = -0.003 mJ/m$^2$ and $D_{yx}$ = -0.009 mJ/m$^2$. $D_{xy}$ and $D_{yx}$ have the same sign indicating opposite spin spiral chiralities along the $x$ and $y$ directions, and, therefore, their combination is expected to be favorable for magnetic antiskyrmions.

We find, however, that the obtained parameters for the polar CrI$_3$ bilayer do not support any magnetic quasiparticles due to a relatively large MAE and a relatively small DMI. This can be seen from our theoretical analysis of the size and shape of skyrmions (antiskyrmions) given in Supporting Information. We find that an (anti)skyrmion has an elliptical shape with a radius that can be estimated as follows:

$$R = \pi D \sqrt{\frac{A'}{16A'K^2 - \pi^2 K D^2}}, \quad (5)$$

where $D = \frac{1}{2}(\alpha|D_{xy}| + |D_{yx}|)$, $A' = \frac{1}{2}(1 + \alpha^2)A$, and $\alpha$ is the ratio of the minor to major axes of the ellipse. For not too strong anisotropy of the DMI, the latter is given by $\alpha = \sqrt{1 + \epsilon^2}$, where

$$\epsilon^2 \approx \frac{\pi^2 |D_{xy}^2 - D_{yx}^2|}{64AK - \frac{1}{2}\pi^2(|D_{xy}| + |D_{yx}|)^2}. \quad (6)$$

In the limit of the isotropic DMI ($\alpha$ = 1), Eq. (5) is reduced to the result obtained in Ref. [67]. Using Eq. (5), where for simplicity we assume $\alpha$ = 1, we obtain the antiskyrmion size of 0.06 Å, which is too small to exist. According to Eq. (5), to realize a sizeable antiskyrmion, a weaker MAE and/or stronger DMI are necessary.

**DFT Results for a Mn-doped CrI$_3$ bilayer.** We propose to use substitutional doping of CrI$_3$ with Mn as a means to decrease MAE. MnI$_3$ is a vdW transition metal trihalide with a structure similar to CrI$_3$. It exhibits a strong magnetic anisotropy in the easy plane [68], which opposes the easy axis anisotropy of CrI$_3$. Our calculations predict that MAE of a MnI$_3$ bilayer in the $A\overline{B}$ stacking configuration is –5.9 meV per Mn atom, compared to 0.67 meV per Cr atom in a polar CrI$_3$ bilayer. Additionally, the lattice constant of MnI$_3$, as determined in our calculations, is 6.73 Å, which is close to the lattice constant of CrI$_3$ of 6.89 Å. Therefore, substitutional Mn doping of CrI$_3$ is feasible experimentally and can be used to make the magnetic anisotropy sufficiently small at a specific mixing ratio.

To calculate the electronic structure and MAE of Cr$_{1-x}$Mn$_x$I$_3$ as a function of Mn doping concentration $x$, we use virtual crystal approximation (VCA). Within VCA, the Cr$_{1-x}$Mn$_x$ substitutional disorder is treated assuming that the Cr-site position in CrI$_3$ is occupied by a fictitious atom with atomic number $24(1 - x) + 25x$, so that at $x$ = 1, the Mn-doped CrI$_3$ corresponds to pure MnI$_3$. With increasing the Mn doping concentration $x$, the electronic structure of Cr$_{1-x}$Mn$_x$I$_3$ changes due to additional valence electrons that occupy the majority-spin $e_g$ band. As a result, the Fermi energy ($E_F$) is pushed to the conduction band, and the band gap separating the $e_g$ and $t_{2g}$ states gets reduced (Fig. 3(c)) at a moderate $x$ and disappears at $x = 1$ (Fig. S1).

As expected and evident from Figure 5(a), an increase of the Mn doping concentration leads to a decrease of MAE per magnetic atom ($K$). At the doping range from $x = 0$ to $x = 0.12$, the anisotropy changes linearly with $x$, nearly reproducing the trend which is expected from the linear interpolation of $K(x)$ from $x = 0$ (CrI$_3$) to $x = 1$ (MnI$_3$) (the blue dashed line in Fig. 5(a)). At $x = 0.12$, we obtain a sufficiently small $K = 0.03$ meV. Therefore, we choose Cr$_{0.88}$Mn$_{0.12}$I$_3$ as a promising candidate to find antiskyrmions.

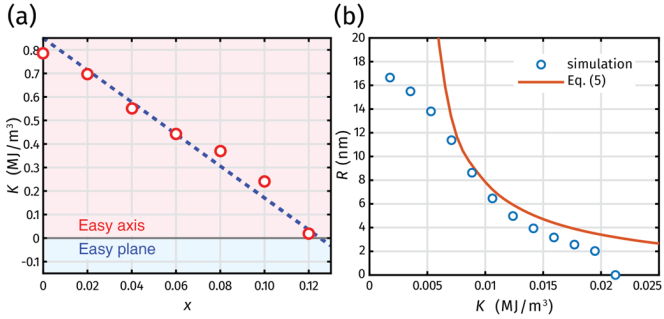

**Figure 5** (a) Calculated magnetic anisotropy $K$ of a Cr$_{1-x}$Mn$_x$I$_3$ polar bilayer as a function of Mn doping $x$ (red circles). The blue dashed line represents a linear interpolation of $K$ between pure CrI$_3$ and MnI$_3$. (b) Antiskyrmion radius $R$ in a Cr$_{0.88}$Mn$_{0.12}$I$_3$ polar bilayer as a function of $K$ obtained from spin-dynamics simulations (blue dots) and estimated using Eq. (5) (red line).

We note that the reduction of MAE with Mn doping can be explained in terms of the imbalance in the population of different $p$ orbitals of the I atoms. It is known that nonmetallic I atoms control the MAE in CrI$_3$ due to the strong spin-orbit coupling associated with them. The sizable out-of-plane magnetic anisotropy of CrI$_3$ can be understood in terms of the second-order perturbation theory by the large contribution of the matrix elements of the orbital momentum between the unoccupied $p_x$ ($p_y$) and occupied $p_y$ ($p_x$) states, while transitions to (from) the $p_z$ states contribute to the in-plane anisotropy [69]. Due to the strong hybridization between the unoccupied transition-metal $e_g$ states and I-$p$ states, population of the former with Mn doping also leads to population of the latter. However, there is a disbalance in the population of the $p_z$ and $p_x$ ($p_y$) orbitals, due to the $p_z$ states lying at a lower energy, as a result of their weaker bonding with the transition-metal $e_g$ states (inset in Fig. S1). A larger population of the $p_z$ states with doping leads to transitions between the occupied $p_z$ states and unoccupied $p_x$ ($p_y$) states, contributing to the in-plane anisotropy.

The calculated magnetic parameters are listed in Table 1. We see that with Mn doping of $x = 0.12$, in addition to significant reduction of magnetic anisotropy, there is an enhancement of the Cr magnetic moment compared to no doping case. This is expected due to the additional majority-spin $e_g$ electrons brought by Mn doping. In addition, we observe that the intralayer exchange coupling $J_{intra}$ and the associated magnetic stiffness $A$ supporting ferromagnetic ordering are increased with the doping. This is because the doped $e_g$ electrons enhance the superexchange interaction [70]. This result is in agreement with the previous calculations suggesting the enhancement of the ferromagnetic exchange with electron doping of CrI$_3$ [71].

We find that the DMI in the Cr$_{0.88}$Mn$_{0.12}$I$_3$ polar bilayer is significantly enhanced compared to CrI$_3$, as seen from the comparison of Figures 4(c) and 4(b). As evident from Table 1, in Cr$_{0.88}$Mn$_{0.12}$I$_3$, the absolute values of $\vec{d}_1$ and $\vec{d}_2$ are 0.08 meV and 0.13 meV, respectively. According to Eq. (3), the DMI tensor is represented by two non-vanishing components: $D_{xy}$ = −0.067 meV and $D_{yx}$ = −0.087 meV, which are an order of magnitude larger than the respective values for CrI$_3$ (Table 1). Similar to the ferromagnetic superexchange interaction, this increase in DMI is driven by a stronger Cr-I bonding that is enhanced by the doped $e_g$ electrons. Both components of the DMI are negative and thus are expected to support antiskyrmions.

**Spin-Dynamics Simulations.** To explore the emergence of antiskyrmions in a polar Cr$_{1-x}$Mn$_x$I$_3$ bilayer, we perform atomistic spin-dynamics simulations using the Landau-Lifshitz-Gilbert (LLG) equation

$$\frac{\partial \vec{m}_i}{\partial t} = -\gamma(\vec{m}_i \times \vec{H}_i) + \alpha_G \left( \vec{m}_i \times \frac{\partial \vec{m}_i}{\partial t} \right). \quad (7)$$

Here $\alpha_G$ is the Gilbert damping constant, $\gamma$ is the gyromagnetic ratio, $\vec{m}_i = \frac{\vec{M}_i}{|\vec{M}_i|}$ is the unit magnetization vector for each sublattice with the magnetization $\vec{M}_i$. The magnetic field $\vec{H}_i = -\frac{1}{\mu}\frac{\partial H}{\partial \vec{m}_i}$ is determined by the spin Hamiltonian:

$$H = -\sum_{i \neq j} J_{ij} \vec{m}_i \cdot \vec{m}_j + \sum_{i \neq j} \vec{d}_{ij} \cdot (\vec{m}_i \times \vec{m}_j) - K \sum_i (\hat{n}_i \cdot \vec{m}_i)^2 \quad (8)$$

where $J_{ij}$ is the exchange coupling, $\vec{d}_{ij}$ is the DMI vector, $K$ is the magnetic anisotropy energy per Cr atom, and $\hat{n}_i$ is the direction of the easy axis. Here, for simplicity, we ignore the magnetic-dipole interactions and the Kitaev-type anisotropic exchange coupling [72] based on the arguments articulated in Discussion and Supporting Information.

In spin-dynamics simulations, we consider all magnetic parameters to be the same as those given in Table 1 for Cr$_{0.88}$Mn$_{0.12}$I$_3$, except magnetic anisotropy $K$ which we assume to be variable in a small window between the calculated value $K = 0.04$ MJ/m$^3$ (0.03 meV/atom) and $K = 0$. This approach is justified by the fact that a change is needed only by a fraction of a percent in Mn concentration to make the anisotropy zero. In this small range of the doping concentrations $x$, other magnetic parameters remain almost intact.

Our spin-dynamics simulations indicate that an antiskyrmion emerges at $K < 0.02$ MJ/m$^2$. At $K = 0.018$ MJ/m$^2$, our results predict that an antiskyrmion has about 2.6 nm radius. Its shape and spin texture are shown in Fig. 6 (a) for polarization of the bilayer pointing upward. It is seen that the antiskyrmion has the spin texture with Néel-type spin twists of opposite chirality along the $x$ and $y$ directions and Bloch-type spin twists of opposite chirality along the two diagonals.

In Figure 5 (b), we plot the calculated antiskyrmion radius as a function of magnetic anisotropy for the parameters of a $Cr_{0.88}Mn_{0.12}I_3$ polar bilayer. It is seen that in the region of $K$ from 0.007 to 0.02 MJ/m$^2$, the antiskyrmion radius (blue dots in Fig. 5(b)) decreases as anisotropy increases, complying with the prediction of Eq. (5) (red line in Fig, 5(b)). For $K$ > 0.02 MJ/m$^2$, an antiskyrmion disappears due to its size being reduced to the lattice constant making it unstable. For small $K$, Eq. (6) predicts divergence of the antiskyrmion radius. We do not see this in our modeling due to the limited simulation cell size and periodic boundary conditions.

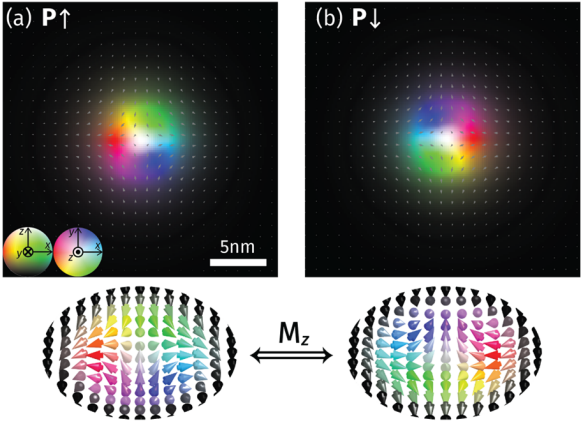

**Figure 6** (**a, b**) Simulated antiskyrmion in polar bilayer $Cr_{0.88}Mn_{0.12}I_3$ with $P\uparrow$ (a) and $P\downarrow$ (b) for $K$ = 0.018 MJ/m$^2$. Polarization switching is equivalent to mirror symmetry $M_z$ transformation of the spin texture, resulting in its 90° rotation.

As was noted above, an anisotropic DMI with $|D_{xy}| \neq |D_{yx}|$ is expected to produce elliptical antiskyrmions with the aspect ratio $\alpha = b/a \neq 1$, where $a$ and $b$ are the semimajor and semiminor axes of the ellipse, respectively. We do observe a deviation of the antiskyrmions from the spherical shape in our simulations. By fitting the polar angle variation of the spin profiles along $x$ and $y$ axes with the function [67,73]:

$$\Theta(r) = 2\arctan\left[\frac{\sinh(R/w)}{\sinh(r/w)}\right], \quad (9)$$

we obtain $a$ = 2.61 nm (semimajor $y$ axis) and $b$ = 2.56 nm (semiminor $x$ axis) for the antiskyrmion in Fig. 6. This implies that the aspect ratio is $\alpha \approx 0.98$, which agrees well with our estimate given by Eq. (6). It is notable that the aspect ratio does not simply scale with the ratio of $D_{xy}$ and $D_{yx}$, which is $D_{xy}/D_{yx} \approx 0.77$ in our case, but is a more complex function of magnetic parameters $D_{xy}$, $D_{yx}$, $K$, and $A$, as follows from Eq. (6) (see Supporting Information for further details).

Importantly, the antiskyrmion spin texture can be controlled by the ferroelectric polarization orientation (Fig. 6). The $P\uparrow$ and $P\downarrow$ structures of a $Cr_{0.88}Mn_{0.12}I_3$ polar bilayer are related by mirror symmetry $M_z$ transformation. This transformation changes the sign of DMI resulting in switching spin chirality of the Néel- and Bloch-type twists (Fig. 6(b)). The associated spin texture is transformed by flipping the in-plane component of spin which is equivalent to the 90-degree rotation of the antiskyrmion. This functionality may be useful for applications in nonvolatile spintronics and quantum computing [74].

## DISCUSSION

Our spin-dynamics simulations do not take into account the magnetic-dipole interactions. It is known that these interactions break the circular symmetry of the antiskyrmion energy affecting their stability and shape [28]. While skyrmions show a homochiral Néel spin texture, antiskyrmions exhibit partly Néel and partly Bloch spin rotations. The latter do not produce magnetic charges making the antiskyrmions energetically more stable. Also, the tendency to favor Bloch rotations over the Néel rotations distorts the spherical (elliptical) shape of antiskyrmions and facilitates the formation of square-shape antiskyrmions, which have been observed experimentally [25-27]. While these effects are interesting, they are expected to emerge at antiskyrmion sizes well above 100 nm and should not influence much the results of our spin-dynamics simulations. Also, dipolar interactions can stabilize larger scale skyrmion-antiskyrmion pairs [75].

As was shown in our studies, MAE is the critical parameter which controls the antiskyrmion formation. Its reduction in $CrI_3$ was required to provide conditions for an antiskyrmion to occur. One could argue that the reduced magnetic anisotropy is expected for $CrCl_3$ and $CrBr_3$, as compared to $CrI_3$, due to a weaker spin-orbit coupling [69]. However, using these materials in a polar bilayer will likely be unfavorable for antiskyrmions because of the reduced DMI which is scaled with spin-orbit coupling. Another possibility to reduce the magnetic anisotropy is to fabricate I-deficient $CrI_3$ which is expected to provide $e_g$-electron doping and thus reduce the anisotropy. However, it is not obvious if the required stoichiometry of the $CrI_{3-x}$ with a relatively large number of I vacancies can be achieved experimentally. The proposed Mn doping of $CrI_3$ seems to be the most viable option to reduce the perpendicular magnetic anisotropy and enhance the DMI.

Forming a moiré superlattice from a twisted $A\overline{A}$-stacked bilayer is an interesting possibility that deserves to be further investigated. It has been recently predicted that a twisted $CrX_3$ (X = I, Br, Cl) bilayer exhibits an effective moiré exchange field resulting in a wide range of noncollinear magnetic phases that can be stabilized as a function of the twist angle in the presence of a non-zero DMI [76]. Making such a twisted bilayer polar using the approach proposed in this work creates conditions for a finite DMI. In this case, however, the DMI is expected to vary along the 2D system plane, which adds more complexity.

Overall, our work demonstrates that polar layer stacking of vdW 2D ferromagnets is an efficient method to generate and modulate DMI. Due to broken space inversion and in-plane rotation symmetries, a polar stacking-engineered bilayer provides necessary conditions for the emergence of magnetic quasiparticles, such as antiskyrmions. Cation doping can be used to tune the magnetic characteristics of the system to achieve the required magnetocrystalline anisotropy and DMI. The spin texture of antiskyrmions can be controlled by ferroelectric polarization in a non-volatile fashion. Such functionality may be interesting for programmable skyrmion-based memories and logic.

Finally, it is worth noting that our proposed approach of using polar stacking as a means to generate the anisotropic DMI is not limited to the specific system but could be applicable to other 2D vdW ferromagnets. Additionally, polar layer stacking that breaks inversion and in-plane rotational symmetries makes it possible to generate other magnetic quasiparticles such as bimerons.

## SUMMARY

In summary, based on first-principles DFT calculations, we have predicted the anisotropic DMI interaction in a $CrI_3$ bilayer engineered by polar layer stacking. Substitutional Mn doping was proposed to reduce the magnetocrystalline anisotropy of pristine $CrI_3$ to observe magnetic quasiparticles. Optimal conditions for the emergence of antiskyrmions were found in a polar $Cr_{0.88}Mn_{0.12}I_3$ bilayer, where a sizable DMI was predicted. Based on atomistic spin-dynamics modelling, the formation of antiskyrmions was demonstrated, with the size tunable by Mn doping. 90° rotation of the spin texture of magnetic antiskyrmions was predicted with reversal of ferroelectric polarization of the bilayer. This possibility of the nonvolatile control of magnetic antiskyrmions in Mn doped $CrI_3$ by an applied electric field is promising for the potential device application of skyrmionic systems.

## ACKNOWLEDGMENTS

This work was primarily supported by the grant number DE-SC0023140 funded by the U.S. Department of Energy, Office of Science, Basic Energy Sciences (K. H., E. Y. T). E. S. and A. A. K. acknowledge support of the EPSCoR RII Track-1 program (NSF Award OIA-2044049) for performing the theoretical analysis of the (anti)skyrmion shape and size and the effect of the Kitaev interaction. Computations were performed at the University of Nebraska Holland Computing Center.